\documentclass[twocolumn,showpacs,showkeys,aps,floatfix]{revtex4-1}
\usepackage{graphicx} 
\usepackage{float}
\usepackage{color}

\begin{document}

\title{The efficient spin injector scheme based on Heusler materials.}

\author{Stanislav~Chadov$^{1}$, Tanja~Graf$^{1}$, Kristina~Chadova$^{2}$, Xuefang~Dai$^{1}$, Frederick~Casper$^{1}$, Gerhard
  H.~Fecher$^{1}$, Claudia~Felser$^{1}$}
\affiliation{$^1$Institut f\"{u}r Anorganische und Analytische Chemie,  Johannes Gutenberg - Universtit\"{a}t,  55099 Mainz,
             Germany \\  $^2$Dept. Chemie und Biochemie, Ludwig Maximilians -  Universit\"at, 81377 M\"unchen, Germany
}

\email{chadov@uni-mainz.de}



\begin{abstract}

We present the rational design scheme intended to provide the 
stable high spin-polarization at the interfaces of the magneto-resistive
junctions by fulfilling the criteria of structural and chemical compatibilities at the interface.
This can be realized by joining the semiconducting and half-metallic
Heusler materials  with similar structures. The present first-principal 
 calculations verify that interface remains half-metallic 
if the nearest interface layers effectively form a stable Heusler material with the  properties 
intermediate between the surrounding bulk parts. This leads to a simple
 rule for selecting the proper combinations.

\end{abstract}

\maketitle 


New  spintronic devices as magneto-resistive (MR)  random access memory
or  read-out  heads  of hard disk  drives~\cite{Zuitic2004,  Coey2002}
require materials exhibiting thermally stable high degree of the
spin-polarization needed for the efficient spin
injection~\cite{deGroot1983, WG01}.  Rich source are the  half-metals
(HM) provided the family of  Co$_2$-based    Heusler    materials
\cite{Kubler1983,Galanakis2002}.  Nowadays they  are the  key candidates
for  the  tunneling  (TMR)  and  giant  magneto-resistance  (GMR)  
devices, with  two HM leads sandwiching a  semiconductor (SC), insulator
or non-magnetic metal spacer layer.  The amazingly high TMR ratios of
570\,\%  at 2~K  were reported for  the system with  HM electrode  of
Co$_2$MnSi with  AlO$_x$  barrier~\cite{Sakuraba2006}. However,
these  values are substantially reduced  at  higher temperatures by the
magnon scattering~\cite{Lezaic2006}.  Besides thermal activation of
magnons other sources for spin depolarization are the various
non-stoichiometries~\cite{KQL+11}, interfacial atomic disorder~\cite{Attema2006}, oxidation, 
spin scattering on defects~\cite{Schmalhorst2004, Telling2006} and even
the intrinsic electron correlation effects~\cite{IKL07}. 

Most of the significant depolarizing mechanisms are directly or indirectly caused by
mechanical factors,  i.\,e. by the lattice mismatch
of HM and SC materials.  By improving the quality of the interface,
e.\,g. by choosing the materials with well matching lattice constants, 
these mechanisms can be substantially
reduced. For example, utilizing MgO as a barrier material and
Co$_2$FeAl$_{0.5}$Si$_{0.5}$ as an electrode, the TMR ratios of
175\,\% could be achieved at room temperature by
now~\cite{Tezuka2006_1}.  On the other hand, even for the well-matching 
interfaces at low temperatures 
the severe problems could be caused by the localized  interface states which couple to
the bulk states of the HM and contribute to the transport
\cite{Mavropoulos2005,KHQF09}.  For example, the significant loss of the
spin-polarization occurs on Co$_2$MnGe/GaAs and Co$_2$CrAl/GaAs
interfaces~\cite{Picozzi2003,NYS06} as well as on the free surfaces.  
Among the rare exclusions is the surface of the Mn-terminated 
Co$_2$MnSi thin film where the HM state is preserved due to the strong 
surface-subsurface coupling~\cite{Hashemifar2005}. In this context the 
importance of studies focusing on  half-metallic
properties of the interface is crucial for designing new efficient 
spin-injecting devices. 

The extremely wide range of electronic properties and rather similar
geometry exhibited by the Heusler family provide the straightforward way
to construct the whole spintronic device by using only  Heusler building blocks. For example, the first-principle calculations~\cite{KHQF09} for
 Co$_2$CrAl(Si)/Cu$_2$CrAl(001) Heusler GMR juntions predicted the spin-polarization of about 80\,\%.
In the following we propose the systematic scheme to search for such proper pairs of
Heusler materials for TMR/GMR junctions and justify it by the first-principle band structure calculations.


The suitable combinations can be derived from the same parent material.
By making various mixtures one can produce the series of new Heusler materials
with smoothly varying electronic properties ranging from half-metallic, magnetic 
to semiconducting and non-magnetic. Much helpful in such design is the 
so-called Slater-Pauling rule~\cite{Slater1936,Pauling1938}  which
states the linear dependency of the unit cell magnetic moment as a
function of the valence electrons number.

\begin{figure}
\centering
\includegraphics[width=1.0\linewidth]{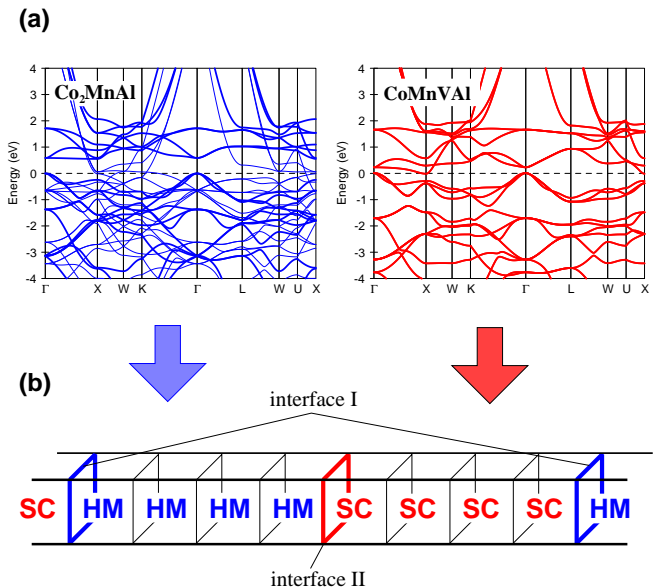}
\caption{ (color online) (a) Calculated  bulk band
  structures of the typical candidate materials: the half-metallic
  ferromagnet ${\rm Co_2MnAl}$ and the nonmagnetic semiconductor ${\rm CoMnVAl}$. 
  In case of ${\rm Co_2MnAl}$ the bands of the gapped
 minority-spin channel are made thicker. The Fermi level is marked by
 the dashed line. (b) Structure of the supercell. Subsections marked
  as HM (half-metal) or SC (semiconductor) represent the complete Heusler
  blocks each containing four atomic layers.}
\label{FIG:JUNCTION}
\end{figure}

We will sketch our idea in details  on example of the well-known 
Heusler Co$_2$MnAl~\cite{BE81} which fulfills the basic requirements
of the efficient spin-injecting material. Band structure
calculations~\cite{KFF07,JYQW08} characterize it as the HM ferromagnet with 
 magnetic moment of 4~$\mu_{\rm B}$ in agreement with experiment. 
Its measured Curie temperature is ${T_{\rm C}=698}$~K~\cite{Webster1971,Jung2009}. 
In order to derive a SC material with a similar lattice 
it is enough to substitute one Co atom by V.  It can be synthesized, 
for example, by 50\,\%-mixing of Mn$_2$VAl~\cite{JVC01} and ${\rm
  Co_2VAl}$~\cite{KCE+10}. The resulting ${\rm CoMnVAl}$ (SC) 
compound  with  24 valence electrons is non-magnetic in agreement with the Slater-Pauling rule. 
The calculated bulk band structures of both Co$_2$MnAl and CoMnVAl are
shown in Figure~\ref{FIG:JUNCTION}\,(a).

To verify which sequence of stacking layers conserves the
half-metallicity  we perform  the corresponding band structure calculations
for the Co$_2$MnAl/CoMnVAl interfaces. In the following example the
stacking direction between ${\rm CoMnVAl}$ and  ${\rm Co_2MnAl}$ is
chosen along the densely packed (001) plane as shown on Figure~\ref{FIG:JUNCTION}\,(b).
Since the supercell contains an integer number of these units, in
general it has no inversion center. For this reason one 
deals with two  nonequivalent interfaces within each supercell. There are
four stacking possibilities along (001) plane which can be paired in two
different supercells. The first one will contain Co-Co/V-Al and Mn-Al/Co-Mn, and the second -
Co-Co/Co-Mn and Mn-Al/Al-V interfaces. Of course, to avoid the interaction between
the interfaces  the supercell must be made as large as possible. 
For this reason all  calculations were performed using the fast {\it ab-initio} LMTO~\cite{PYA} 
method which accurately calculates the band structure within
$\sim1$\,Ry around the Fermi energy. This allows to study the reasonably
large supercell containing 64 atoms arranged within 32 atomic layers. 
The exchange-correlation potential is treated using the
 Vosko-Wilk-Nusair form of the LSDA \cite{VWN80}. Since the spin-orbit
 coupling is sufficiently small in these systems  the  calculations are 
performed in the scalar-relativistic regime.


The optimization of the supercell volume yields the lattice constant
 nearly equal to the average of the optimized  bulk values for the HM
and SC materials which mismatch by about 2\,\%.
The maximal size of the supercell (CoMnVAl)$_4$/(Co$_2$MnAl)$_4$ (64 nonequivalent atoms arranged 
within 32 atomic planes) is already enough to achieve a good agreement
 for the layer-resolved density of states (DOS) at the Fermi energy and magnetic moments of the inner layers with the
 corresponding bulk values. 

\begin{figure}[htdb]
\centering
\includegraphics[width=1.0\linewidth]{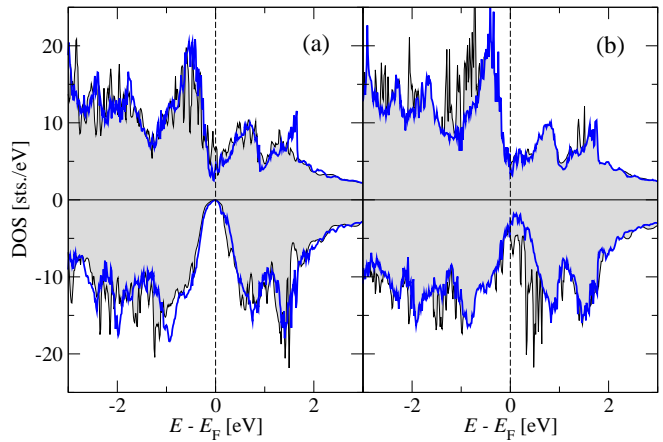}
\caption{(color online) Spin-resolved DOS (positive values correspond to majority-spin
  negative - to minority-spin channel)  of the HM/SC supercells (HM =
  ${\rm Co_2MnAl}$, SC = ${\rm CoMnVAl}$) with Co-Co/V-Al, Mn-Al/Co-Mn (a) and Co-Co/Mn-Co, Mn-Al/Al-V
  (b) interface pairs. The gray shaded area corresponds to the minimal
  size of the supercell (HM)$_1$/(SC)$_1$ (scaled up by factor 4), the
  thick blue curve corresponds to the largest (HM)$_4$/(SC)$_4$ supercell.}
\label{FIG:DOS-TOTAL}
\end{figure}

\begin{figure}
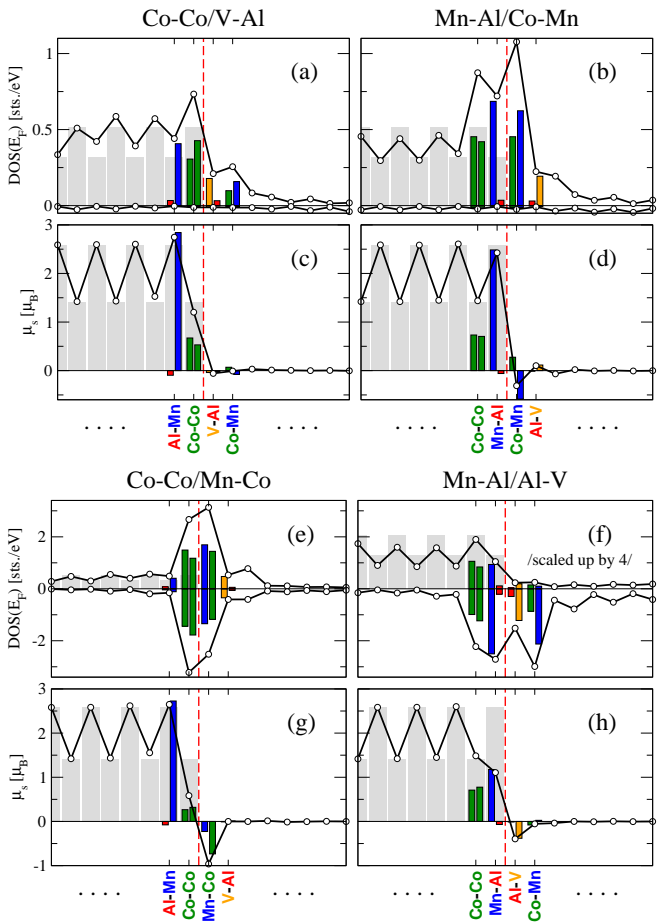

\centering
\includegraphics[width=1.0\linewidth]{fig3a.eps}\\[1ex]
\includegraphics[width=1.0\linewidth]{fig3b.eps}
\caption{(color online) Black solid line (with opened circles) represent the layer-resolved DOS at the
  Fermi energy  with positive values referring to the   majority-spin, negative - to the minority-spin channels (a, b, e, f)
  and the   magnetic moments (c, d, g, h)  calculated for  HM/SC-supercells (HM = Co$_2$MnAl, SC = CoMnVAl). 
  Gray bars show the corresponding values calculated for the bulk HM and SC
  materials.  Colored bars mark the atom-projected contributions within the first nearest and next-nearest interface
  layers (each layer contains two atoms). Vertical red dashed lines mark
  the interface borders.}
\label{FIG:PROFILES}
\end{figure}

As it follows from the spin-resolved DOS curves (Figure~\ref{FIG:DOS-TOTAL}) the spin polarization 
 indeed depends critically on the way of stacking: independently 
on the system size the  half-metallicity is preserved 
for the system with  Co-Co/V-Al and Co-Mn/Mn-Al interfaces 
and  in case of Co-Co/Mn-Co and Mn-Al/Al-V interfaces it is destroyed. 
The substantially higher total energy (by about 3~mRy) of the
supercell with ``destructive''  interfaces indicates their relative instability. In the ``destructive'' case 
the minority-spin DOS at the Fermi level for the largeer (HM)$_4$/(SC)$_4$ supercell is noticeably
lower than for the smaller one, (HM)$_1$/(SC)$_1$. This obviously tells that
the ``destructive'' states originate locally from the interface layers.
This can be viewed in more details by considering the layer-resolved 
DOS($E_{\rm F}$) and the magnetization profiles  (Figure~\ref{FIG:PROFILES}). 

At  Co-Co/V-Al (a), and  Co-Mn/Mn-Al (b)
interfaces, the spin polarization increases. At the same time at
Co-Co/Mn-Co (e)  and Al-V/Mn-Al (f) interfaces it is destroyed.
The reason can be qualitatively understood by comparing  the
 materials effectively formed on the interfaces with their ideal bulk
 equivalents, since the properties of Heuslers in a large extent originate
from the nearest neighbor coupling. 
Indeed, Co-Co/V-Al and Co-Mn/Mn-Al interfaces correspond to the existing 
Heusler compounds with 2~$\mu_{\rm B}$ magnetic moment and high spin
polarization: Co$_2$VAl~\cite{JVC01,JYQW08} and Mn$_2$CoAl~\cite{LDL+08}.
As it follows from Figure~\ref{FIG:PROFILES}, except of the overall
demagnetization,  the magnetic structure of these interfaces 
is rather similar to their bulk equivalents.
Indeed, first of all, both interfaces
are half-metallic. For Co-Co/V-Al interface, the magnetic moments of Co
atoms (0.65 and 0.5~$\mu_{\rm B}$) are about two times 
smaller comparing to the corresponding ${\rm Co_2VAl}$ bulk
material (about 1~$\mu_{\rm B}$).  However, similar to the bulk, they are both
positive and are followed by the non-magnetic V-Al layer.
Similar situation also occurs at Mn-Al/Co-Mn interface: the atomic
moments are approximately $2.5, 0.78$ and $-0.68~\mu_{\rm B}$, comparing to  $3, 1$ and
$-2~\mu_{\rm B} $ for Mn, Co and the second Mn atom respectively.

The other two compounds, ${\rm AlVMnAl}$ and ${\rm CoCoMnCo}$ which
form the ``destructive'' interfaces, are equivalent 
to ${\rm MnVAl_2}$ and ${\rm Co_3Mn}$ which to our knowledge do not
exist in the Heusler structure. Indeed, as we mentioned above, their
 calculated total energies are noticeably higher than for
 ``constructive'' interfaces and they exist mainly due to the coupling
 with the outer layers.


Thus we can conclude that the ``constructive'' interface (preserving the half-metallicity) 
can be formed if the effective interface composition would correspond to
the stable bulk material with the intermediate properties between the 
left- and right-side materials, as in the sequence of ${\rm
  Co_2MnAl}$/${\rm Co_2VAl}$/${\rm CoMnVAl}$ which exhibit the bulk
magnetic moments of 4, 2 and 0~$\mu_{\rm B}$, respectively. The
experimentally suitable  method to obtain the
24-electron SC material would be through a mixture of two stable
HM ferromagnets with  numbers of valence electrons larger and smaller
than 24. 

This situation is rather general. By applying similar first-principle 
analysis we have justified the analogous situation for the series of 
other Co$_2$-based Heusler materials. The pairs of ``constructive'' and ``destructive'' interfaces
were found also for ${\rm Co_2MnZ/CoMnTiZ}$ (Z=Si, Ge, Sn) and
${\rm Co_2FeZ/CoFeTiZ}$ (Z=Al, Ga) (more details can be found in
Ref.~\footnote{C.~Felser, F.~Casper, X.~Dai and G.~Reiss,
  Patent~DE102008046920.3}). 
For the ``constructive'' case the effective
interface compounds will correspond to Co$_2$TiZ  group of half-metallic
ferromagnets with magnetic moments of 1~$\mu_{\rm B}$.

The particularly selected  (001)-orientation of the interfaces is also not
unique. For example, we have  verified that the same rules 
apply for (111)-orientation as well.  This  goes in line with  the
general idea that the most important condition
 is the relative smooth change of properties while going from the
 ferromagnetic half-metallic to a non-magnetic semiconducting Heusler
 system.

\begin{acknowledgments}

Financial support by  DFG Research unit FOR~559 (projects P01, P07), 
the Bundesministerium f\"ur Bildung und Forschung BMBF 
(Project MultiMag), the Graduate School of Excellence MAInZ and 
the DFG/ASPIMATT project (unit 1.2-A) are gratefully acknowledged.

\end{acknowledgments}


\begin{thebibliography}{10}%
\makeatletter
\providecommand \@ifxundefined [1]{%
 \ifx #1\undefined \expandafter \@firstoftwo
 \else \expandafter \@secondoftwo
\fi
}%
\providecommand \@ifnum [1]{%
 \ifnum #1\expandafter \@firstoftwo
 \else \expandafter \@secondoftwo
\fi
}%
\providecommand \enquote [1]{``#1''}%
\providecommand \bibnamefont  [1]{#1}%
\providecommand \bibfnamefont [1]{#1}%
\providecommand \citenamefont [1]{#1}%
\providecommand\href[0]{\@sanitize\@href}%
\providecommand\@href[1]{\endgroup\@@startlink{#1}\endgroup\@@href}%
\providecommand\@@href[1]{#1\@@endlink}%
\providecommand \@sanitize [0]{\begingroup\catcode`\&12\catcode`\#12\relax}%
\@ifxundefined \pdfoutput {\@firstoftwo}{%
 \@ifnum{\z@=\pdfoutput}{\@firstoftwo}{\@secondoftwo}%
}{%
 \providecommand\@@startlink[1]{\leavevmode}%
 \providecommand\@@endlink[0]{}%
}{%
 \providecommand\@@startlink[1]{%
  \leavevmode
  \pdfstartlink
   attr{/Border[0 0 1 ]/H/I/C[0 1 1]}%
   user{/Subtype/Link/A<</Type/Action/S/URI/URI(#1)>>}%
  \relax
 }%
 \providecommand\@@endlink[0]{\pdfendlink}%
}%
\providecommand \url  [0]{\begingroup\@sanitize \@url }%
\providecommand \@url [1]{\endgroup\@href {#1}{\urlprefix}}%
\providecommand \urlprefix [0]{URL }%
\providecommand \Eprint[0]{\href }%
\@ifxundefined \urlstyle {%
  \providecommand \doi [1]{doi:\discretionary{}{}{}#1}%
}{%
  \providecommand \doi [0]{doi:\discretionary{}{}{}\begingroup
  \urlstyle{rm}\Url }%
}%
\providecommand \doibase [0]{http://dx.doi.org/}%
\providecommand \Doi[1]{\href{\doibase#1}}%
\providecommand \bibAnnote [3]{%
  \BibitemShut{#1}%
  \begin{quotation}\noindent
    \textsc{Key:}\ #2\\\textsc{Annotation:}\ #3%
  \end{quotation}%
}%
\providecommand \bibAnnoteFile [2]{%
  \IfFileExists{#2}{\bibAnnote {#1} {#2} {\input{#2}}}{}%
}%
\providecommand \typeout [0]{\immediate \write \m@ne }%
\providecommand \selectlanguage [0]{\@gobble}%
\providecommand \bibinfo [0]{\@secondoftwo}%
\providecommand \bibfield [0]{\@secondoftwo}%
\providecommand \translation [1]{[#1]}%
\providecommand \BibitemOpen[0]{}%
\providecommand \bibitemStop [0]{}%
\providecommand \bibitemNoStop [0]{.\EOS\space}%
\providecommand \EOS [0]{\spacefactor3000\relax}%
\providecommand \BibitemShut [1]{\csname bibitem#1\endcsname}%
\bibitem{Zuitic2004}%
  \BibitemOpen
  \bibfield{author}{%
  \bibinfo {author} {\bibfnamefont{I.}~\bibnamefont{\v{Z}uti\'{c}}}, \bibinfo
  {author} {\bibfnamefont{J.}~\bibnamefont{Fabian}},\ and\ \bibinfo {author}
  {\bibfnamefont{S.~D.}\ \bibnamefont{Sarma}},\ }%
  \bibfield{journal}{%
  \bibinfo {journal} {Rev. Mod. Phys.}\ }%
  \textbf{\bibinfo {volume} {76}},\ \bibinfo {pages} {323} (\bibinfo {year}
  {2004})%
  \bibAnnoteFile{NoStop}{Zuitic2004}%
\bibitem{Coey2002}%
  \BibitemOpen
  \bibfield{author}{%
  \bibinfo {author} {\bibfnamefont{J.~M.~D.}\ \bibnamefont{Coey}}\ and\
  \bibinfo {author} {\bibfnamefont{M.~N.~B.}\ \bibnamefont{M~Venkatesan}},\ }%
  \emph{\bibinfo {title} {Lecture Notes in Physics}},\ \bibinfo {edition}
  {595th}\ ed.\ (\bibinfo {publisher} {Springer, Heidelberg},\ \bibinfo {year}
  {2002})\ p.\ \bibinfo {pages} {377}%
  \bibAnnoteFile{NoStop}{Coey2002}%
\bibitem{deGroot1983}%
  \BibitemOpen
  \bibfield{author}{%
  \bibinfo {author} {\bibfnamefont{R.~A.}\ \bibnamefont{de~Groot}}, \bibinfo
  {author} {\bibfnamefont{F.~M.}\ \bibnamefont{Mueller}}, \bibinfo {author}
  {\bibfnamefont{P.~G.}\ \bibnamefont{van Engen}},\ and\ \bibinfo {author}
  {\bibfnamefont{K.~H.~J.}\ \bibnamefont{Buschow}},\ }%
  \bibfield{journal}{%
  \bibinfo {journal} {Phys. Rev. Lett.}\ }%
  \textbf{\bibinfo {volume} {50}},\ \bibinfo {pages} {2024} (\bibinfo {year}
  {1983})%
  \bibAnnoteFile{NoStop}{deGroot1983}%
\bibitem{WG01}%
  \BibitemOpen
  \bibfield{author}{%
  \bibinfo {author} {\bibfnamefont{G.~A.}\ \bibnamefont{de~Wijs}}\ and\
  \bibinfo {author} {\bibfnamefont{R.~A.}\ \bibnamefont{de~Groot}},\ }%
  \bibfield{journal}{%
  \bibinfo {journal} {Phys. Rev. B}\ }%
  \textbf{\bibinfo {volume} {64}},\ \bibinfo {pages} {020402} (\bibinfo {year}
  {2001})%
  \bibAnnoteFile{NoStop}{WG01}%
\bibitem{Kubler1983}%
  \BibitemOpen
  \bibfield{author}{%
  \bibinfo {author} {\bibfnamefont{J.}~\bibnamefont{K\"{u}bler}}, \bibinfo
  {author} {\bibfnamefont{A.~R.}\ \bibnamefont{Williams}},\ and\ \bibinfo
  {author} {\bibfnamefont{C.~B.}\ \bibnamefont{Sommers}},\ }%
  \bibfield{journal}{%
  \bibinfo {journal} {Phys. Rev. B}\ }%
  \textbf{\bibinfo {volume} {28}},\ \bibinfo {pages} {1745} (\bibinfo {year}
  {1983})%
  \bibAnnoteFile{NoStop}{Kubler1983}%
\bibitem{Galanakis2002}%
  \BibitemOpen
  \bibfield{author}{%
  \bibinfo {author} {\bibfnamefont{I.}~\bibnamefont{Galanakis}}, \bibinfo
  {author} {\bibfnamefont{P.~H.}\ \bibnamefont{Dederichs}},\ and\ \bibinfo
  {author} {\bibfnamefont{N.}~\bibnamefont{Papanikolaou}},\ }%
  \bibfield{journal}{%
  \bibinfo {journal} {Phys. Rev. B}\ }%
  \textbf{\bibinfo {volume} {66}},\ \bibinfo {pages} {174429} (\bibinfo {year}
  {2002})%
  \bibAnnoteFile{NoStop}{Galanakis2002}%
\bibitem{Sakuraba2006}%
  \BibitemOpen
  \bibfield{author}{%
  \bibinfo {author} {\bibfnamefont{Y.}~\bibnamefont{Sakuraba}}, \bibinfo
  {author} {\bibfnamefont{M.}~\bibnamefont{Hattori}}, \bibinfo {author}
  {\bibfnamefont{M.}~\bibnamefont{Oogane}}, \bibinfo {author}
  {\bibfnamefont{Y.}~\bibnamefont{Ando}}, \bibinfo {author}
  {\bibfnamefont{H.}~\bibnamefont{Kato}}, \bibinfo {author}
  {\bibfnamefont{A.}~\bibnamefont{Sakuma}}, \bibinfo {author}
  {\bibfnamefont{T.}~\bibnamefont{Miyazaki}},\ and\ \bibinfo {author}
  {\bibfnamefont{H.}~\bibnamefont{Kubota}},\ }%
  \bibfield{journal}{%
  \bibinfo {journal} {Appl. Phys. Lett.}\ }%
  \textbf{\bibinfo {volume} {88}},\ \bibinfo {pages} {192508} (\bibinfo {year}
  {2006})%
  \bibAnnoteFile{NoStop}{Sakuraba2006}%
\bibitem{Lezaic2006}%
  \BibitemOpen
  \bibfield{author}{%
  \bibinfo {author} {\bibfnamefont{M.}~\bibnamefont{Lezaic}}, \bibinfo {author}
  {\bibfnamefont{P.}~\bibnamefont{Mavropoulos}}, \bibinfo {author}
  {\bibfnamefont{J.}~\bibnamefont{Enkovaara}}, \bibinfo {author}
  {\bibfnamefont{G.}~\bibnamefont{Bihlmayer}},\ and\ \bibinfo {author}
  {\bibfnamefont{S.}~\bibnamefont{Bl\"{u}gel}},\ }%
  \bibfield{journal}{%
  \bibinfo {journal} {Phys. Rev. Lett.}\ }%
  \textbf{\bibinfo {volume} {97}},\ \bibinfo {pages} {026404} (\bibinfo {year}
  {2006})%
  \bibAnnoteFile{NoStop}{Lezaic2006}%
\bibitem{KQL+11}%
  \BibitemOpen
  \bibfield{author}{%
  \bibinfo {author} {\bibfnamefont{V.}~\bibnamefont{Ko}}, \bibinfo {author}
  {\bibfnamefont{J.}~\bibnamefont{Qiu}}, \bibinfo {author}
  {\bibfnamefont{P.}~\bibnamefont{Luo}}, \bibinfo {author}
  {\bibfnamefont{G.~C.}\ \bibnamefont{Han}},\ and\ \bibinfo {author}
  {\bibfnamefont{Y.~P.}\ \bibnamefont{Feng}},\ }%
  \bibfield{journal}{%
  \bibinfo {journal} {J.~Appl.~Phys.}\ }%
  \textbf{\bibinfo {volume} {109}},\ \bibinfo {pages} {07B103} (\bibinfo {year}
  {2011})%
  \bibAnnoteFile{NoStop}{KQL+11}%
\bibitem{Attema2006}%
  \BibitemOpen
  \bibfield{author}{%
  \bibinfo {author} {\bibfnamefont{J.~J.}\ \bibnamefont{Attema}}, \bibinfo
  {author} {\bibfnamefont{G.~A.}\ \bibnamefont{de~Wijs}},\ and\ \bibinfo
  {author} {\bibfnamefont{R.~A.}\ \bibnamefont{de~Groot}},\ }%
  \bibfield{journal}{%
  \bibinfo {journal} {J. Phys. D: Appl. Phys.}\ }%
  \textbf{\bibinfo {volume} {39}},\ \bibinfo {pages} {793} (\bibinfo {year}
  {2006})%
  \bibAnnoteFile{NoStop}{Attema2006}%
\bibitem{Schmalhorst2004}%
  \BibitemOpen
  \bibfield{author}{%
  \bibinfo {author} {\bibfnamefont{J.}~\bibnamefont{Schmalhorst}}, \bibinfo
  {author} {\bibfnamefont{S.}~\bibnamefont{K\"ammerer}}, \bibinfo {author}
  {\bibfnamefont{M.}~\bibnamefont{Sacher}}, \bibinfo {author}
  {\bibfnamefont{G.}~\bibnamefont{Reiss}}, \bibinfo {author}
  {\bibfnamefont{A.}~\bibnamefont{H\"utten}},\ and\ \bibinfo {author}
  {\bibfnamefont{A.}~\bibnamefont{Scholl}},\ }%
  \bibfield{journal}{%
  \bibinfo {journal} {Phys. Rev. B}\ }%
  \textbf{\bibinfo {volume} {70}},\ \bibinfo {pages} {024426} (\bibinfo {year}
  {2004})%
  \bibAnnoteFile{NoStop}{Schmalhorst2004}%
\bibitem{Telling2006}%
  \BibitemOpen
  \bibfield{author}{%
  \bibinfo {author} {\bibfnamefont{N.~D.}\ \bibnamefont{Telling}}, \bibinfo
  {author} {\bibfnamefont{P.~S.}\ \bibnamefont{Keatley}}, \bibinfo {author}
  {\bibfnamefont{G.}~\bibnamefont{van~der Laan}}, \bibinfo {author}
  {\bibfnamefont{R.~J.}\ \bibnamefont{Hicken}}, \bibinfo {author}
  {\bibfnamefont{E.}~\bibnamefont{Arenholz}}, \bibinfo {author}
  {\bibfnamefont{Y.}~\bibnamefont{Sakuraba}}, \bibinfo {author}
  {\bibfnamefont{M.}~\bibnamefont{Oogane}}, \bibinfo {author}
  {\bibfnamefont{Y.}~\bibnamefont{Ando}},\ and\ \bibinfo {author}
  {\bibfnamefont{T.}~\bibnamefont{Miyazaki}},\ }%
  \bibfield{journal}{%
  \bibinfo {journal} {Phys. Rev. B}\ }%
  \textbf{\bibinfo {volume} {74}},\ \bibinfo {pages} {224439} (\bibinfo {year}
  {2006})%
  \bibAnnoteFile{NoStop}{Telling2006}%
\bibitem{IKL07}%
  \BibitemOpen
  \bibfield{author}{%
  \bibinfo {author} {\bibfnamefont{V.~Y.}\ \bibnamefont{Irkhin}}, \bibinfo
  {author} {\bibfnamefont{M.~I.}\ \bibnamefont{Katsnelson}},\ and\ \bibinfo
  {author} {\bibfnamefont{A.~I.}\ \bibnamefont{Lichtenstein}},\ }%
  \bibfield{journal}{%
  \bibinfo {journal} {J. Phys.: Cond. Mat.}\ }%
  \textbf{\bibinfo {volume} {19}},\ \bibinfo {pages} {315201} (\bibinfo {year}
  {2007})%
  \bibAnnoteFile{NoStop}{IKL07}%
\bibitem{Tezuka2006_1}%
  \BibitemOpen
  \bibfield{author}{%
  \bibinfo {author} {\bibfnamefont{N.}~\bibnamefont{Tezuka}}, \bibinfo {author}
  {\bibfnamefont{N.}~\bibnamefont{Ikeda}}, \bibinfo {author}
  {\bibfnamefont{S.}~\bibnamefont{Sugimoto}},\ and\ \bibinfo {author}
  {\bibfnamefont{K.}~\bibnamefont{Inomata}},\ }%
  \bibfield{journal}{%
  \bibinfo {journal} {Appl. Phys. Lett.}\ }%
  \textbf{\bibinfo {volume} {89}},\ \bibinfo {pages} {252508} (\bibinfo {year}
  {2006})%
  \bibAnnoteFile{NoStop}{Tezuka2006_1}%
\bibitem{Mavropoulos2005}%
  \BibitemOpen
  \bibfield{author}{%
  \bibinfo {author} {\bibfnamefont{P.}~\bibnamefont{Mavropoulos}}, \bibinfo
  {author} {\bibfnamefont{M.}~\bibnamefont{Lezaic}},\ and\ \bibinfo {author}
  {\bibfnamefont{S.}~\bibnamefont{Bl\"ugel}},\ }%
  \bibfield{journal}{%
  \bibinfo {journal} {Phys. Rev. B}\ }%
  \textbf{\bibinfo {volume} {72}},\ \bibinfo {pages} {174428} (\bibinfo {year}
  {2005})%
  \bibAnnoteFile{NoStop}{Mavropoulos2005}%
\bibitem{KHQF09}%
  \BibitemOpen
  \bibfield{author}{%
  \bibinfo {author} {\bibfnamefont{V.}~\bibnamefont{Ko}}, \bibinfo {author}
  {\bibfnamefont{G.}~\bibnamefont{Han}}, \bibinfo {author}
  {\bibfnamefont{J.}~\bibnamefont{Qiu}},\ and\ \bibinfo {author}
  {\bibfnamefont{Y.~P.}\ \bibnamefont{Feng}},\ }%
  \bibfield{journal}{%
  \bibinfo {journal} {J.~Appl.~Phys.}\ }%
  \textbf{\bibinfo {volume} {95}},\ \bibinfo {pages} {202502} (\bibinfo {year}
  {2009})%
  \bibAnnoteFile{NoStop}{KHQF09}%
\bibitem{Picozzi2003}%
  \BibitemOpen
  \bibfield{author}{%
  \bibinfo {author} {\bibfnamefont{S.}~\bibnamefont{Picozzi}}, \bibinfo
  {author} {\bibfnamefont{A.}~\bibnamefont{Continenza}},\ and\ \bibinfo
  {author} {\bibfnamefont{A.~J.}\ \bibnamefont{Freeman}},\ }%
  \bibfield{journal}{%
  \bibinfo {journal} {J. Phys. Chem. Solids}\ }%
  \textbf{\bibinfo {volume} {64}},\ \bibinfo {pages} {1697} (\bibinfo {year}
  {2003})%
  \bibAnnoteFile{NoStop}{Picozzi2003}%
\bibitem{NYS06}%
  \BibitemOpen
  \bibfield{author}{%
  \bibinfo {author} {\bibfnamefont{K.}~\bibnamefont{Nagao}}, \bibinfo {author}
  {\bibfnamefont{Y.}~\bibnamefont{Miura}},\ and\ \bibinfo {author}
  {\bibfnamefont{M.}~\bibnamefont{Shirai}},\ }%
  \bibfield{journal}{%
  \bibinfo {journal} {Phys.~Rev.~B}\ }%
  \textbf{\bibinfo {volume} {73}},\ \bibinfo {pages} {104447} (\bibinfo {year}
  {2006})%
  \bibAnnoteFile{NoStop}{NYS06}%
\bibitem{Hashemifar2005}%
  \BibitemOpen
  \bibfield{author}{%
  \bibinfo {author} {\bibfnamefont{S.~J.}\ \bibnamefont{Hashemifar}}, \bibinfo
  {author} {\bibfnamefont{P.}~\bibnamefont{Kratzer}},\ and\ \bibinfo {author}
  {\bibfnamefont{M.}~\bibnamefont{Scheffler}},\ }%
  \bibfield{journal}{%
  \bibinfo {journal} {Phys. Rev. Lett.}\ }%
  \textbf{\bibinfo {volume} {94}},\ \bibinfo {pages} {096402} (\bibinfo {year}
  {2005})%
  \bibAnnoteFile{NoStop}{Hashemifar2005}%
\bibitem{Slater1936}%
  \BibitemOpen
  \bibfield{author}{%
  \bibinfo {author} {\bibfnamefont{J.~C.}\ \bibnamefont{Slater}},\ }%
  \bibfield{journal}{%
  \bibinfo {journal} {Phys. Rev.}\ }%
  \textbf{\bibinfo {volume} {49}},\ \bibinfo {pages} {931} (\bibinfo {year}
  {1936})%
  \bibAnnoteFile{NoStop}{Slater1936}%
\bibitem{Pauling1938}%
  \BibitemOpen
  \bibfield{author}{%
  \bibinfo {author} {\bibfnamefont{L.}~\bibnamefont{Pauling}},\ }%
  \bibfield{journal}{%
  \bibinfo {journal} {Phys. Rev.}\ }%
  \textbf{\bibinfo {volume} {54}},\ \bibinfo {pages} {899} (\bibinfo {year}
  {1938})%
  \bibAnnoteFile{NoStop}{Pauling1938}%
\bibitem{BE81}%
  \BibitemOpen
  \bibfield{author}{%
  \bibinfo {author} {\bibfnamefont{K.~H.~J.}\ \bibnamefont{Buschow}}\ and\
  \bibinfo {author} {\bibfnamefont{P.~G.}\ \bibnamefont{van Engen}},\ }%
  \bibfield{journal}{%
  \bibinfo {journal} {J. Magn. Magn. Mat.}\ }%
  \textbf{\bibinfo {volume} {25}},\ \bibinfo {pages} {90} (\bibinfo {year}
  {1981})%
  \bibAnnoteFile{NoStop}{BE81}%
\bibitem{KFF07}%
  \BibitemOpen
  \bibfield{author}{%
  \bibinfo {author} {\bibfnamefont{J.}~\bibnamefont{K\"{u}bler}}, \bibinfo
  {author} {\bibfnamefont{G.~H.}\ \bibnamefont{Fecher}},\ and\ \bibinfo
  {author} {\bibfnamefont{C.}~\bibnamefont{Felser}},\ }%
  \bibfield{journal}{%
  \bibinfo {journal} {Phys. Rev. B}\ }%
  \textbf{\bibinfo {volume} {76}},\ \bibinfo {pages} {024414} (\bibinfo {year}
  {2007})%
  \bibAnnoteFile{NoStop}{KFF07}%
\bibitem{JYQW08}%
  \BibitemOpen
  \bibfield{author}{%
  \bibinfo {author} {\bibfnamefont{X.}~\bibnamefont{Jia}}, \bibinfo {author}
  {\bibfnamefont{W.}~\bibnamefont{Yang}}, \bibinfo {author}
  {\bibfnamefont{M.}~\bibnamefont{Qin}},\ and\ \bibinfo {author}
  {\bibfnamefont{L.}~\bibnamefont{Wang}},\ }%
  \bibfield{journal}{%
  \bibinfo {journal} {J. Phys. D: Appl. Phys.}\ }%
  \textbf{\bibinfo {volume} {41}},\ \bibinfo {pages} {085004} (\bibinfo {year}
  {2008})%
  \bibAnnoteFile{NoStop}{JYQW08}%
\bibitem{Webster1971}%
  \BibitemOpen
  \bibfield{author}{%
  \bibinfo {author} {\bibfnamefont{P.~J.}\ \bibnamefont{Webster}},\ }%
  \bibfield{journal}{%
  \bibinfo {journal} {J. Phys. Chem. Solids}\ }%
  \textbf{\bibinfo {volume} {32}},\ \bibinfo {pages} {1221} (\bibinfo {year}
  {1971})%
  \bibAnnoteFile{NoStop}{Webster1971}%
\bibitem{Jung2009}%
  \BibitemOpen
  \bibfield{author}{%
  \bibinfo {author} {\bibfnamefont{V.}~\bibnamefont{Jung}}, \bibinfo {author}
  {\bibfnamefont{G.~H.}\ \bibnamefont{Fecher}}, \bibinfo {author}
  {\bibfnamefont{B.}~\bibnamefont{Balke}}, \bibinfo {author}
  {\bibfnamefont{V.}~\bibnamefont{Ksenofontov}},\ and\ \bibinfo {author}
  {\bibfnamefont{C.}~\bibnamefont{Felser}},\ }%
  \bibfield{journal}{%
  \bibinfo {journal} {J. Phys. D: Appl. Phys.}\ }%
  \textbf{\bibinfo {volume} {42}},\ \bibinfo {pages} {084007} (\bibinfo {year}
  {2009})%
  \bibAnnoteFile{NoStop}{Jung2009}%
\bibitem{JVC01}%
  \BibitemOpen
  \bibfield{author}{%
  \bibinfo {author} {\bibfnamefont{C.}~\bibnamefont{Jiang}}, \bibinfo {author}
  {\bibfnamefont{M.}~\bibnamefont{Venkatesan}},\ and\ \bibinfo {author}
  {\bibfnamefont{J.~M.~D.}\ \bibnamefont{Coey}},\ }%
  \bibfield{journal}{%
  \bibinfo {journal} {Solid State Commun.}\ }%
  \textbf{\bibinfo {volume} {118}},\ \bibinfo {pages} {513} (\bibinfo {year}
  {2001})%
  \bibAnnoteFile{NoStop}{JVC01}%
\bibitem{KCE+10}%
  \BibitemOpen
  \bibfield{author}{%
  \bibinfo {author} {\bibfnamefont{T.}~\bibnamefont{Kanomata}}, \bibinfo
  {author} {\bibfnamefont{Y.}~\bibnamefont{Chieda}}, \bibinfo {author}
  {\bibfnamefont{K.}~\bibnamefont{Endo}}, \bibinfo {author}
  {\bibfnamefont{H.}~\bibnamefont{Okada}}, \bibinfo {author}
  {\bibfnamefont{M.}~\bibnamefont{Nagasako}}, \bibinfo {author}
  {\bibfnamefont{K.}~\bibnamefont{Kobayashi}}, \bibinfo {author}
  {\bibfnamefont{R.}~\bibnamefont{Kainuma}}, \bibinfo {author}
  {\bibfnamefont{R.~Y.}\ \bibnamefont{Umetsu}}, \bibinfo {author}
  {\bibfnamefont{H.}~\bibnamefont{Takahashi}}, \bibinfo {author}
  {\bibfnamefont{Y.}~\bibnamefont{Furutani}}, \bibinfo {author}
  {\bibfnamefont{H.}~\bibnamefont{Nishihara}}, \bibinfo {author}
  {\bibfnamefont{K.}~\bibnamefont{Abe}}, \bibinfo {author}
  {\bibfnamefont{Y.}~\bibnamefont{Miura}},\ and\ \bibinfo {author}
  {\bibfnamefont{M.}~\bibnamefont{Shirai}},\ }%
  \bibfield{journal}{%
  \bibinfo {journal} {Phys.~Rev.~B}\ }%
  \textbf{\bibinfo {volume} {82}},\ \bibinfo {pages} {144415} (\bibinfo {year}
  {2010})%
  \bibAnnoteFile{NoStop}{KCE+10}%
\bibitem{PYA}%
  \BibitemOpen
  \bibfield{author}{%
  \bibinfo {author} {\bibfnamefont{A.}~\bibnamefont{Perlov}}, \bibinfo {author}
  {\bibfnamefont{A.}~\bibnamefont{Yaresko}},\ and\ \bibinfo {author}
  {\bibfnamefont{V.}~\bibnamefont{Antonov}},\ }%
  \enquote{\bibinfo {title} {\textit{Spin-polarized Relativistic Linear
  Muffin-tin Orbitals Package for Electronic Structure Calculations,
  PY-LMTO.}}.}\ %
  \bibAnnoteFile{NoStop}{PYA}%
\bibitem{VWN80}%
  \BibitemOpen
  \bibfield{author}{%
  \bibinfo {author} {\bibfnamefont{S.~H.}\ \bibnamefont{Vosko}}, \bibinfo
  {author} {\bibfnamefont{L.}~\bibnamefont{Wilk}},\ and\ \bibinfo {author}
  {\bibfnamefont{M.}~\bibnamefont{Nusair}},\ }%
  \bibfield{journal}{%
  \bibinfo {journal} {Canad.~J.~Phys.}\ }%
  \textbf{\bibinfo {volume} {58}},\ \bibinfo {pages} {1200} (\bibinfo {year}
  {1980})%
  \bibAnnoteFile{NoStop}{VWN80}%
\bibitem{LDL+08}%
  \BibitemOpen
  \bibfield{author}{%
  \bibinfo {author} {\bibfnamefont{G.~D.}\ \bibnamefont{Liu}}, \bibinfo
  {author} {\bibfnamefont{X.~F.}\ \bibnamefont{Dai}}, \bibinfo {author}
  {\bibfnamefont{H.~Y.}\ \bibnamefont{Liu}}, \bibinfo {author}
  {\bibfnamefont{J.~L.}\ \bibnamefont{Chen}}, \bibinfo {author}
  {\bibfnamefont{Y.~X.}\ \bibnamefont{Li}}, \bibinfo {author}
  {\bibfnamefont{G.}~\bibnamefont{Xiao}},\ and\ \bibinfo {author}
  {\bibfnamefont{G.~H.}\ \bibnamefont{Wu}},\ }%
  \bibfield{journal}{%
  \bibinfo {journal} {Phys. Rev. B}\ }%
  \textbf{\bibinfo {volume} {77}},\ \bibinfo {pages} {014424} (\bibinfo {year}
  {2008})%
  \bibAnnoteFile{NoStop}{LDL+08}%
\bibitem{Note1}%
  \BibitemOpen
  \bibinfo {note} {C.~Felser, F.~Casper, X.~Dai and G.~Reiss,
  Patent~DE102008046920.3}%
  \bibAnnoteFile{NoStop}{Note1}%
\end{thebibliography}
\end{document}